# Bringing the welcome home: One section's efforts at incorporating AAPT's diversity and inclusion practices

Bree Barnett Dreyfuss[1] and David Marasco[2]
[1]Amador Valley High School, Pleasanton, CA 94566
[2]Foothill College, Los Altos Hills, CA 94062

While AAPT and many other physics organizations have been introducing a series of effective practices around diversity and inclusion at the national level in recent years, it was wondered if these were being adopted at the local level. It is hoped that section members and Section Representatives will decide to further expand the actions of national leadership to make their own section meetings more inclusive.  In order to assess if this was in fact the case, a survey on diversity practices that have been used at the national level and can be implemented at the section level was sent to AAPT's Section Reps mailing list in the spring of 2018, with a follow-up survey in winter 2020.  Feedback in both cycles suggested that a guide for section leadership would be useful. The Northern California/Nevada section has made progress in implementing some of the effective practices from the national meetings into our local section meetings, we share these efforts in the hope that they assist our fellow sections.

**Practices involving our membership, keynote speakers and panelists:**
**Intentional recruitment for diversity of invited speakers[1,2,3,4]**
When inviting keynote speakers, the Northern California/ Nevada section has actively recruited speakers from a variety of cultural backgrounds, genders, disciplines, and career levels.  Of the twenty-five invited speakers in the last eight years, twelve have been women and four have identified as part of other traditionally marginalized groups.  At our fall 2016 meeting we were fortunate that Mary Gaillard, the first female tenured physics professor at UC Berkeley, participated in a panel of women in physics.  Recognizing that wisdom and insight is present in members of varying levels of experience, we have invited speakers that are grad students, just starting on the tenure track, tenured professors, Nobel Laureates, nonfiction authors, government and industry scientists, noneducators and civil rights activists. Students at the host school are encouraged to present their projects and posters to our membership as a way to practice for presentations along their career pathways.

**Intentional planning of at least one "diversity" topic/activity at each meeting**

A prevailing attitude of our leadership team in recent years has been "I teach students" rather than just "I teach physics." As such, in our section the regular offering of diversity topics for our meetings has been as much organic as it has been intentional.

While our officers and membership possess some content knowledge in diversity and inclusion, it has been beneficial to reach out to other resources.  In 2017, a representative from the local section of the American Civil Liberties Union spoke to our membership about both our rights and those of the students we teach in the current political climate. If the local host is a college or university, the Ethnic and Gender Studies Departments can be strong allies.  Framing a speaker request in terms of "would you give an hour or two on a Saturday to help other teachers better understand these important topics?" is a good way to start the conversation.  People who might not want to develop a talk may be willing to serve on a panel.  We have found that it is powerful to have student voices on these panels, but it is first important to establish their sense of safety, both emotionally and physically. We have seen that if a person with authority who has a good relationship with the student is a member or moderator of the panel, this can be sufficient. Additional steps considered for the safety of the panelists included reading a statement setting norms for questions from the audience, and/or having questions submitted via 3x5 cards and filtered through the moderator. Another benefit of a 3x5 card, or an electronic anonymized system, is that people can ask questions that they may be uncomfortable to ask verbally.  It is helpful to discuss these options with students, or any other member of a vulnerable population who sits on a panel.  At the City College of San Francisco in Fall of 2017 Professor Ardel Haefele-Thomas, chair of the Lesbian, Gay, Bisexual, Transgender and Queer Studies Department, was part of a panel with a pair of their students that identified as nonbinary or transgender. The panelists had a chance to share their personal educational experiences including which of their teachers' practices made them feel the most comfortable, accepted and seen. As many of our members were not well-versed in issues facing transgendered people, it provided an opportunity for them to hear from students and an educator about how they can best create a safe and welcoming environment for all their students to learn physics.

**Intentional diversity in the images used in recruitment/advertising materials[4]**
Twice a year over a thousand postcards go to all physics teachers that have attended a section meeting in the last five years, as well as to each high school, university and two year college in our region. Most postcards announcing our meetings include pictures of recent meetings, members and keynote speakers. We have purposely chosen images that represent our diverse membership and invited speakers. When diversity or equity-centered programming is planned we will also highlight them in our announcements. Since brand new teachers through emeritus faculty attend our meetings we purposely choose images and statements that reflect that people of all experience levels are welcome (Figures 1,2, and 3).

**Purposeful Practice in Section Leadership**
Recognizing that there is often a large turnover in leadership as members advance within their institutions, national organizations or in their own lives, ten years ago the section leadership actively recruited new officers. It was a conscious choice to recruit while there were still more experienced people involved to mentor them. Due to the efforts of past officers, when measured broadly across ethnicity, gender, gender identity, and sexual orientation, section leadership is far more diverse than our membership. We have continued this model of encouraging others to step forward in leadership before we have multiple officer vacancies, with an eye to diversity. This has brought forward new ideas and program innovations as our collective perspective is broadened by our different life experiences and viewpoints.[1]

While granting formal awards and recognitions has not been a big part of our section's culture, we do occasionally honor a person at the local level, and we do send nominations for AAPT Fellows on a regular basis. We had been using the lens of "lifetime achievement" as the main deciding factor, but realized that this had the effect of selecting almost exclusively from the previous generation, which does not represent the whole of our current membership. Moving forward, we are looking at honoring members for their current contributions.

## Dependent-care grants

In 2017 our section requested and was granted an AAPT Section Mini-Grant[5] to provide support to people who have family-care obligations during our meetings. While the original target was people who had stopped attending due to the addition of infants or young children to their families[6,7], we have expanded the grant to include any dependent care needs. The grant funds now cover any meeting attendee, regardless of previous attendance status, or if their family responsibilities involved other loved ones rather than children such as elderly parents. We offered a $75 stipend to attendees who needed dependent care for one person, and $150 for more than one, which is tailored to the cost of living in Northern California. The AAPT Section Mini-Grant funding allowed our section to run this program for two years, and the application process was administered through a short Google Form[8]. We have awarded over $1000 in funds over the last two years to meeting attendees, and AAPT has granted a one-time renewal. We warn that there should be the expectation of no-shows, as we are intentionally reaching out to a population that has regular barriers to attendance, this should be expected. Our experience is that this is true roughly one-third of the time. When we receive a last minute decline we encourage members to re-apply and try to attend the next meeting. The funds are granted at the meeting and those allocated to someone that cannot show are returned to the account. Based upon our application form, we have mainly served single parents and people whose partners work weekends. We have been told that these grants are what makes attendance possible for these members, and several have been able to attend multiple meetings.

## Use of an Event Participation Code of Conduct[4,9,10]

We adopted an Event Participation Code of Conduct[11] at our Fall 2019 meeting. The Code of Conduct was based on AAPT's, but was kept to one page to encourage people to read the policy. This length was possible as we could refer back to definitions and other useful language in AAPT's original document[12]. The Code of Conduct was posted to our website under meeting information, emailed to all pre-registered participants and a poster was placed near the registration table for walk-in registrants. While our section has not had any reported incidents in the current officers' memory, we adopted a Code of Conduct to reinforce expectations of professional conduct at our meetings, and to further create an atmosphere of safety to minoritized populations. We plan to iterate our Code, with improvements such as a clearer reporting system for possible incidents.

**ADA comment box for online registration and technology to support people with disabilities**[13]

Our section has relied upon online registration for over five years.  We have included the question "Generally, our events are held at schools with basic ADA standards. If you have any accessibility needs, let us know here and we'll work to accommodate them."  The Americans with Disabilities Act (ADA) "gives civil rights protections to individuals with disabilities similar to those provided to individuals on the basis of race, color, sex, national origin, age, and religion," with the intent of giving people with disabilities the same access and opportunites as everyone else.[14] Needs we have seen most often have been food sensitivity issues, and easy access to bathrooms.  We do have an unwritten policy that an assistant or caretaker of a member can attend our meetings without paying for registration (only paying for lunch), but we recognize we should do a better job of publicizing this practice.

An area for growth in our section includes requiring the use of microphones and PA systems, as well as increased use of QR codes or short links posters and slide presentations. Prior to our most recent meeting, our section had not required the use of microphones, although they are often available, which is especially shameful as one of our prominent long-time members has had trouble hearing for many years. This member is known for sitting in the front row at each meeting and has had to ask speakers to repeat themselves on numerous occasions. This member found the microphone to be unhelpful when implemented at our most recent meeting, so we plan to explore transcribing software.  The auto captioning in Microsoft PowerPoint365 is promising.  Requiring access to a functional audio system will be part of our future facility requirements for a meeting location, and our section is investigating options for purchasing a suitable portable audio system for when we are in venues that cannot provide a proper sound system. Like many practices in the classroom, people will conform to norms, once the expectation of microphone use was established, all members used it.  Asking members to include a QR code or short link on their presentations that will take them to an online version further ensures access to the materials for all members if they have trouble viewing or hearing the presentation in the meeting conditions. Links to presentation slides are now sent out to participants after meetings.  Our best efforts to ensure that all members can hear presenters and see their presentations does include equitable access to technology. Supplying a variety of common laptop to projector connectors will ensure presenters can project anything they wish and limit the number of presenters that have to present without their planned visuals.

After the global pandemic of 2020 forced many educators to implement distance learning, many of us became more aware of the many online platforms for recording and sharing materials and videos. We hope to add recorded and closed captioned videos of meetings in the future.

**Meeting and Facility practices:**

**Gender pronoun stickers[1,4]**

Our section requested AAPT's gender pronoun materials following the winter 2017 meeting in Atlanta, and implemented them at our next meeting.  We printed AAPT's explanation document as a poster, placed it next to registration, and printed a selection of pronoun stickers.  We have made this toolkit available to other AAPT sections via our website[15]. Initially, some explanation on the purpose behind the practice was necessary as some members were not aware of the need.[16]  After the practice had been introduced to our section, we made preferred pronouns an optional part of our online registration process, and printed pronouns directly on name tags for our members. Nearly 60% of the registrants at our most recent meeting registered with preferred pronouns. It is important to not require this, as there are safety issues surrounding asking a marginalized person to out themselves to an organization that has not yet earned that level of trust.  Those that register the day of the event were encouraged to write their pronouns on their nametags.  We have recently introduced 1.25" buttons with traditional male/female pronouns, they/them/theirs and "Ask me my pronouns" as options.  This is seen as preferable to printing/writing on name tags. A member can simply remove a button if they no longer feel safe, perhaps while outside the main meeting space and elsewhere at the facility, which is a better option than having to remove their name tag. It is also possible that a member may feel more comfortable removing their button, even if what leads to this does not rise to the level of a Code of Conduct violation. The buttons can also be used at other events where pronoun stickers are not available, and additionally serve as advertising for our section.

**Gender-neutral bathrooms[1]**

California has required gender-neutral bathroom options in educational settings since the early part of this decade.  While gender-neutral bathrooms have been in use at most of our meeting sites, their level of accessibility tends to be site specific. In some locations a gender-neutral faculty bathroom is available, in others the only bathrooms in the building are student bathrooms and gendered. Typically the availability issue has arisen when we visited sites with older buildings. As our leadership team continues to standardize our host site responsibilities, we intend to include an easily accessible gender-neutral bathroom as one of our facility requirements and include a clearly marked map that shows their location.

**Lactation room**

Years ago a member that needed to feed or pump breastmilk for their child was left to find their own space in which to do so during a meeting, if they attended at all. We have now added a lactation room[4,7,17] to our facility requirements and advertised on our announcement postcards the fact that one will be available to attendees. At a minimum, the space should include a place to sit, an electric outlet and a lockable door. If the room itself does not include a sink then it should be in close proximity to a bathroom. Access to a refrigerator is also recommended but most people storing milk or formula come prepared with their own cold storage. The room has been used by at least one member in four of the five meetings since we've established this practice.

## Conclusion

Some of the actions detailed above have been in place for nearly a decade, others have been more recently implemented. Have they led to a difference in who attends our meetings? In order to compare demographics, attendee rosters from our spring 2007, spring 2008, and fall 2008 meetings were reconstructed. At those three meetings we averaged roughly 11 women per meeting, accounting for 17% of our attendees. For the corresponding meetings ten years later, the average was 32 women for 33% of our attendees. Acknowledging that attempting to identify people's ethnicity is problematic, especially via names on a roster or from a group photo, anecdotally we have also become more diverse in terms of race. We would like to thank with deep gratitude Paul "Pablo" Robinson and Dennis Buckley for providing and continuing to provide the leadership that has made NCNAAPT a thriving, active chapter. We appreciate their initial work to make our section more inclusive and welcoming to more physics teachers. We recognize that our progress is incomplete and ongoing.

## Acknowledgements

This work benefited from insightful feedback from Beth Cunningham, Val Monticue, and the reviewers. We would like to recognize Dennis Buckley for reconstituting the older attendance rosters and maintaining them for our section for several decades, and also David Sturm for his support in thinking about how diversity and inclusion practices could be implemented at the local level.

Figures 1 and 2: Examples of announcement postcards sent to all physics departments and members in Northern California and Nevada. Often the keynote speakers are pictured so care is taken to present all equally, given biases that can surround title, experience, and demographics.
Photos of members from previous meetings are selected to highlight the diversity of our membership, and welcoming language is used to encourage attendance from across our membership.

Figure 3: An enlargement of one of the group photos. Group photos are often a part of our postcards, we make a point of taking a group photo to promote a sense of belonging, and when time permits distribute 4x6s prior to the end of each meeting.

# NCNAAPT invites YOU to our Fall 2018 Conference & Meeting

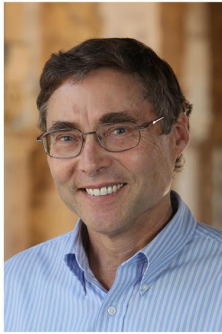

## November 9th and 10th

### Keynote talks by:

**Carl Wieman**
Nobel Prize Winner & co-founder of PhET

**AND**

**Tracy Van Houten**
JPL Engineer & political activist

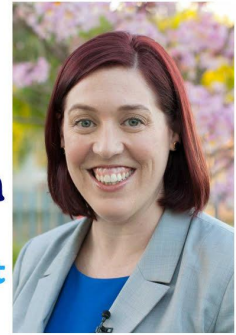

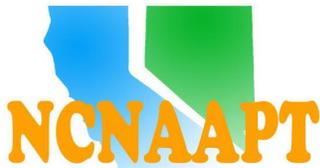

Saturday conference at Bellarmine College Prepatory in San Jose

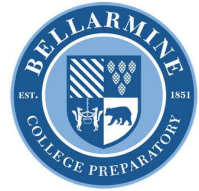

@NCNAAPT

#NCNAAPT

New to teaching? Veteran teacher? High School? College? All are welcome!

## By Physics Teachers. For Physics Teachers.

# NCNAAPT invites YOU to our Spring 2017 Conference & Meeting

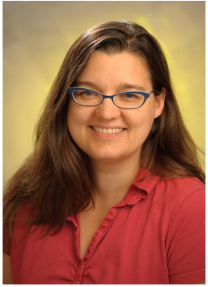

**April 28th & April 29th**

**Keynote talk by Jessie Dotson of NASA**

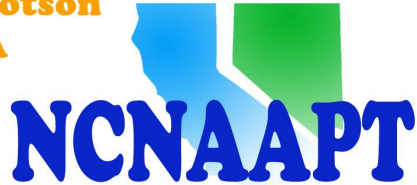

**NCNAAPT**

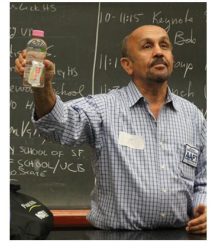
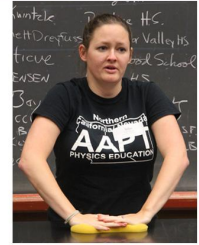

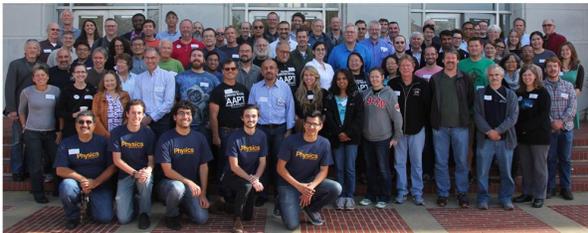

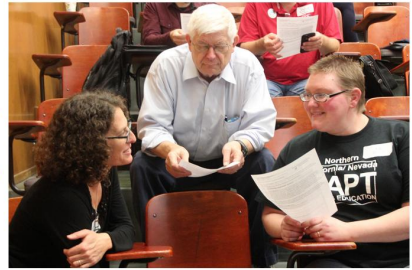

🐦 **@NCNAAPT**

**f #NCNAAPT**

**NCNAAPT.org**

**New to teaching? Veteran teacher? High School? College? All are welcome!**

# By Physics Teachers. For Physics Teachers.

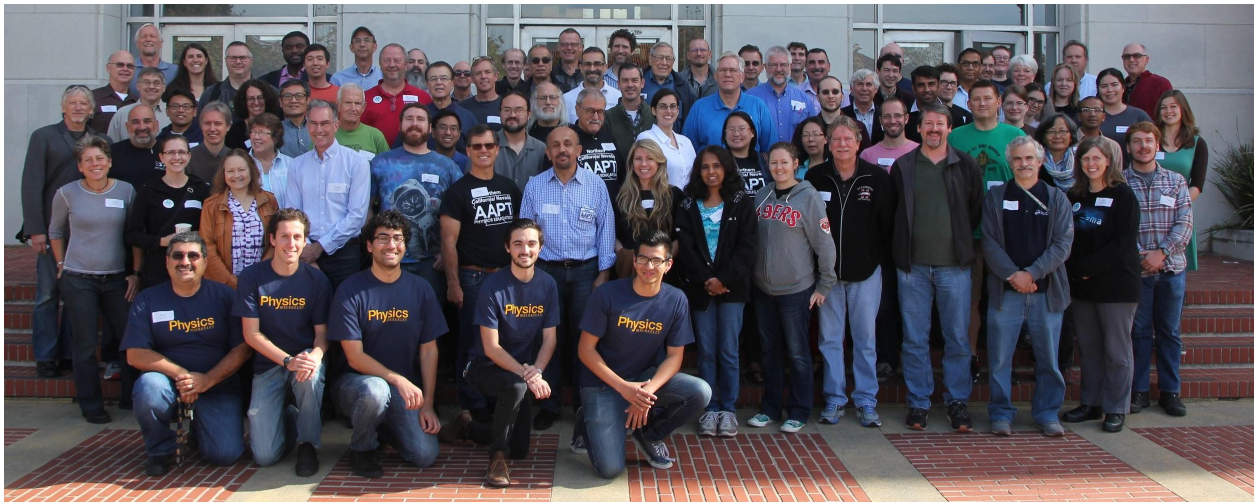